\documentstyle[aps]{revtex}
\begin{document}
\title{Primordial Magnetic Fields induced by Cosmological Particle Creation}
\author{Esteban A. Calzetta\thanks{%
Electronic address: calzetta@df.uba.ar}$^{1,2}$, Alejandra Kandus$^{1,2}$%
\thanks{%
Electronic address: kandus@df.uba.ar} and Francisco D. Mazzitelli$^{1,2}$%
\thanks{%
Electronic address: fmazzi@df.uba.ar}}
\address{$^1$Instituto de Astronom\'\i a y F\'\i sica del Espacio, \\
c.c. 67, suc 28 (1428) Buenos Aires Argentina and \\
$^2$Departamento de F\'\i sica, Facultad de Ciencias Exactas y Naturales, \\
Ciudad Universitaria (1428) Buenos Aires, Argentina}
\maketitle

\begin{abstract}
We study the primordial magnetic field generated by stochastic currents
produced by scalar charged particles created at the beginning of the
radiation dominated epoch. We find that for the mass range $%
10^{-6}GeV\lesssim m\lesssim 10^2GeV$, a field of sufficient intensity to
seed different mechanisms of galactic magnetic field generation, while still
consistent with observational and theoretical constraints, is created
coherently over a galactic scale.
\end{abstract}

\section{Introduction}

At present there exists huge observational evidence about the presence of
magnetic fields throughout the Universe: our own galaxy is endowed with a
homogeneous magnetic field $B\simeq 3\times 10^{-6}$ $Gauss$ and similar
field intensity is detected in high redshift galaxies \cite{kronberg} and in
damped Lyman alpha clouds \cite{wolfe}.

The origin of these fields is still unclear. Research is performed mainly
along the idea of a cosmological mechanism: a seed primordial field that
would be further amplified by protogalactic collapse and differential
rotation or a nonlinear dynamo\cite{zeldovich}. Several mechanisms have been
proposed to explain the origin of the seed field. It has been suggested that
a primordial field may be produced during the inflationary period if
conformal invariance is broken \cite{widrow}, \cite{dolgov}. In
string-inspired models, the coupling between the electromagnetic field and
the dilaton breaks conformal invariance and may produce the seed field\cite
{ratra}. Gauge invariant couplings between the electromagnetic field and the
space-time curvature also break conformal invariance but produce, in
general, an uninterestingly small seed field \cite{federico}. Other
mechanisms are based on a first order cosmological phase transition \cite
{olinto} and on the existence of topological defects \cite{others}.

In this paper we propose a new mechanism for primordial field generation in
the early Universe, based on stochastic currents generated by particle
creation of scalar charged species \cite{birrel}. We assume the presence
during Inflation of a charged, minimally coupled, scalar field in its
invariant vacuum state \cite{allen}.When the transition to a radiation
dominated Universe takes place, quantum creation of charged particles
occurs. We assume that the field mass is smaller than the vacuum energy
density during Inflation, $H$, and therefore can consider the transition
from that period to Radiation dominance as instantaneous.

As the number of positive charged species is the same as the number of
negative charged ones (there is no physical reason why it should not be so),
the mean electric current is zero. Nevertheless quantum fluctuations around
the mean give rise to a non-vanishing current. We compute the rms amplitude
of these fluctuations and use it as the source term in the equation for the
magnetic field. We must stress the fact that the field must be a scalar
minimally coupled one: it is straightforward to check that with a massive
conformally coupled scalar field very few particles are created and
therefore a very weak magnetic field is created. For spinorial fields we
show in the Appendix that, due to the conformal invariance of massless
fermions, the number of particles created is very small and consequently the
magnetic field produced is extremely weak.

\section{Calculation of the magnetic field}

As the process of magnetic field generation that we are studying takes place
after inflation, there is no loss of generality if we work in a spatially
flat Universe. In conformal time, $d\tau =dt/a(t),$ we have $g_{\mu \nu
}=a^2(\tau )\eta _{\mu \nu }$, $\eta _{\mu \nu }$ being the minkowskian
metric tensor. The canonical scalar field, which we assume to be massive and
minimally coupled, is written as $\varphi =\phi /a$.

Part of this process takes place before the electroweak transition. Then
part of the created photons are a combination of of the isospin and
hypercharge bosons, where the coefficients are respectively the $\sin $ and $%
\cos $ of the Weinberg angle $\theta _w$ \cite{ratra}, \cite{quigg}. We will
consider the magnetic field generated by only the hypercharge sector and in
this sense the figures obtained constitute a lower bound to the effective
intensity of the field. The amplitude of the electromagnetic field would be
smaller than the pure $U(1)$ boson by a factor $\cos \theta _w$, but
recalling that the electroweak coupling constant is $g=q/\cos \theta _w$, we
obtain the same amplitude for the created electromagnetic field. The
magnetic field is then defined from the spatial components of the field
tensor, $B_i=(1/2)\varepsilon _{ijk}F_{jk}$, where $F_{\mu \nu }=\partial
_\mu A_\nu -\partial _\nu A_\mu $, $A_\mu $ being the vector potential in
the Lorentz gauge $A_{;\mu }^\mu =0.$

The equation for $\vec B$ reads:

\begin{equation}
\left[ \frac{\partial ^2}{\partial \tau ^2}-\nabla ^2+\sigma \left( \tau
\right) \frac \partial {\partial \tau }\right] \vec B=\vec \nabla \times 
\vec j  \label{ad}
\end{equation}
where $\sigma \left( \tau \right) $ is a time-dependent conductivity of the
plasma and $\vec j$ represents the spacelike components of the electric
current which for the scalar field is $j_\mu =iq\left( \phi ^{*}\partial
_\mu \phi -\phi \partial _\mu \phi ^{*}\right) +2q^2A_\mu \phi ^{*}\phi $.

The basic point in our analysis is that while the expectation value of the
current vanishes, there are quantum and statistical fluctuations which build
up a nonvanishing root mean square value. The source in these equations can
be phenomenologically substituted by a stochastic, Gaussian current\cite
{beilok}. Our goal is to estimate the magnetic field produced by this
stochastic source. Under the assumption that the source is Gaussian, its
statistics are fixed once the two point source-source correlation function
is given. Since the radiated field will be weak in any case, as a first
approximation we can compute this correlation function in the absence of a
macroscopic electromagnetic field.

As $\left\langle \vec j\right\rangle =0$, the only contribution to the
electric current will be due to the quantum fluctuations of the fields. In
order to study if the quantum fluctuations of the massive fields produce a
non zero electric current we calculate the two point correlation function

\begin{equation}
N_{ii^{\prime }}\left[ (\tau ,\vec r),(\tau ^{\prime },\vec r^{\prime
})\right] \equiv \left\langle \left\{ \left( \vec \nabla \times \vec j%
\right) _i(\tau ,\vec r),\left( \vec \nabla \times \vec j\right) _{i^{\prime
}}(\tau ^{\prime },\vec r^{\prime })\right\} \right\rangle  \label{ag}
\end{equation}

The scalar field can be written in terms of two real fields according to $%
\phi (\tau ,\vec r)=\left\{ \phi _1(\tau ,\vec r)+i\phi _2(\tau ,\vec r%
)\right\} /\sqrt{2}$ so that the spacelike part of the current $j_\mu $, at
vanishing e.m. field reads $\vec j=q\left\{ \phi _2\vec \nabla \phi _1-\phi
_1\vec \nabla \phi _2\right\} $ and 
\begin{equation}
\vec \nabla \times \vec j=2q\left( \vec \nabla \phi _2\right) \times \left( 
\vec \nabla \phi _1\right)  \label{af}
\end{equation}

If we consider that $\phi _1$ and $\phi _2$ commute, we can write 
\begin{equation}
N_{ii^{\prime }}\left[ (\tau ,\vec r),(\tau ^{\prime },\vec r^{\prime
})\right] =4q^2\epsilon _{ijl}\epsilon _{i^{\prime }j^{\prime }l^{\prime
}}\left\{ \partial _{jj^{\prime }}^2G^{+}\partial _{ll^{\prime
}}^2G^{+}+\partial _{jj^{\prime }}^2G^{-}\partial _{ll^{\prime
}}^2G^{-}\right\}  \label{ah}
\end{equation}
where we have introduced positive and negative frequency propagators $%
G^{+}\equiv \left\langle \phi _i(\tau ,\vec r)\phi _i(\tau ^{\prime },\vec r%
^{\prime })\right\rangle $ and $G^{-}\equiv \left\langle \phi _i(\tau
^{\prime },\vec r^{\prime })\phi _i(\tau ,\vec r)\right\rangle $. Writing
the propagators in terms of their Fourier components we get 
\begin{equation}
N_{ii^{\prime }}\left[ (\tau ,\vec r),(\tau ^{\prime },\vec r^{\prime
})\right] =4q^2\int \frac{d\vec \kappa \;d\vec \kappa ^{\prime }}{(2\pi )^3}%
\left( \vec \kappa \times \vec \kappa ^{\prime }\right) _i\left( \vec \kappa
\times \vec \kappa ^{\prime }\right) _{i^{\prime }}e^{i(\vec \kappa +\vec 
\kappa ^{\prime }).(\vec r-\vec r^{\prime })}G_\kappa ^{+}(\tau ,\tau
^{\prime })G_{\kappa ^{\prime }}^{+}(\tau ,\tau ^{\prime })+(\tau ,\vec r%
\longleftrightarrow \tau ^{\prime },\vec r^{\prime })  \label{ai}
\end{equation}
Expanding the fields as $\phi =\left( 2\pi \right) ^{-3/2}\int d^3\kappa
\;\phi _\kappa (\tau )e^{i\vec \kappa .\vec r}+h.c.$ the Fourier transform
of the propagators is given by $G_\kappa ^{+}(\tau ,\tau ^{\prime })=\phi
_\kappa (\tau )\phi _\kappa ^{*}(\tau ^{\prime })=G_\kappa ^{-}(\tau
^{\prime },\tau )$.

Our next task is the evaluation of the scalar field modes. In the absence of
electromagnetic fields, the scalar field satisfies the Klein - Gordon
equation

\begin{equation}
\left[ \frac{\partial ^2}{\partial \tau ^2}+{\kappa}^2+m^2a^2(\tau )- \frac{%
\ddot a\left( \tau \right) }{a\left( \tau \right) }\right] \phi _\kappa
(\tau )=0  \label{aii}
\end{equation}

We assume instantaneous reheating at $\tau =0$, 
\begin{equation}
a(\tau )=\left\{ 
\begin{array}{cc}
\frac 1{H(\tau _0-\tau )} & Inflation \\ 
&  \\ 
\frac 1{H\tau _0}\left( 1+\frac \tau {\tau _0}\right) & Radiation
\end{array}
\right.  \label{aj}
\end{equation}
and normalize the scale factor by taking $H\tau _0=1$. In terms of the
dimensionless variables $y_0=H\tau $ and $k=\kappa /H$ the modes in the
inflationary epoch read 
\begin{equation}
\phi _k^I(y_0)=\frac{\sqrt{\pi }}2\sqrt{1-y_0}H_\nu ^{(1)}\left[
k(1-y_0)\right] ,\;\nu =\frac 32\sqrt{1-\frac 49\frac{m^2}{H^2}}  \label{ak}
\end{equation}
For the radiation dominated era we write 
\begin{equation}
\phi _k^{WKB}(y_0)=\alpha _kf_k^0+\beta _kf_k^{0*}  \label{al}
\end{equation}
where $f_k^0$ is the WKB solution

\[
f_k^0\left( y_0\right) \sim \frac{e^{-i\Omega _k(y_0)}}{\sqrt{2\omega _k(y_0)%
}} 
\]
where $\omega _k(y_0)=\sqrt{k^2+\frac{m^2}{H^2}a(y_0)^2}$, $\Omega
_k(y_0)=\int^{y_0}dy_0^{\prime }\omega _k(y_0^{\prime })$. $\alpha _k$ and $%
\beta _k$ are the Bogolubov coefficients connecting the modes (\ref{ak})
with the WKB basis (\ref{al}) at $y_0=0$. In the limit $k\ll m/H\ll 1$ they
are given by 
\begin{equation}
\beta _k\simeq -\alpha _k\simeq -i\frac{\Gamma \left( 3/2\right) }4\sqrt{%
\frac{\pi H}m}\frac 1{k^{3/2}}  \label{am}
\end{equation}
The non zero value of $\beta _k$ reflects particle creation from the
gravitational field, rather than from the decay of the inflaton field.
Indeed the occupation number for long wavelength modes diverges as $k^{-3}$,
much in excess of the $k^{-1}$ Rayleigh - Jeans tail of the thermal spectrum
produced by the reheating process. This excess noise results in an
enhancement of the magnetic field over and above the equilibrium
fluctuations. When the given scale enters the horizon, particle -
antiparticle annihilation becomes efficient, and the extra noise disappears.
It is worth mentioning that the logarithmic divergence in the total number
of particles created up to $k_{\max }$ can be removed with a suitable
infrared cut-off, say the mode corresponding to the horizon $\sim 10^{-26}$.
The energy density associated to these particles, $\rho =mH^3\int^{k_{\max
}}d^3k\left| \beta _k\right| ^2$, after a cutoff is imposed is $\sim
10^{-16} $ smaller than the one of the CMBR.

To give a quantitative estimate of the magnetic field amplitude, let us come
back to the evaluation of the two point correlation function (\ref{ag}),
that can be written as $N_{ii^{\prime }}=N_{ii^{\prime }}^0+N_{ii^{\prime
}}^1+N_{ii^{\prime }}^2$. The first term is the noise that would be present
in the absence of particle creation during reheating and does not concern us
here. The other terms are the contribution due to the particles created at
the beginning of the radiation era and are given by 
\begin{equation}
N_{ii^{\prime }}^1\left[ \left( \tau ,\vec r\right) ,\left( \tau ^{\prime },%
\vec r^{\prime }\right) \right] \rightarrow 4q^2H^8\int \frac{d\vec k\,d\vec 
k^{\prime }}{\left( 2\pi \right) ^3}\left( \vec k\times \vec k^{\prime
}\right) _i\left( \vec k\times \vec k^{\prime }\right) _{i^{\prime
}}e^{i\left( \vec k+\vec k^{\prime }\right) .\left( \vec y-\vec y^{\prime
}\right) }G_{1k}^0(y_0,y_0^{\prime })\delta G_{1k^{\prime }}(y_0,y_0^{\prime
})  \label{ao}
\end{equation}
\begin{equation}
N_{ii^{\prime }}^2\left[ \left( \tau ,\vec r\right) ,\left( \tau ^{\prime },%
\vec r^{\prime }\right) \right] \rightarrow q^2H^8\int \frac{d\vec k\,d\vec k%
^{\prime }}{\left( 2\pi \right) ^3}\left( \vec k\times \vec k^{\prime
}\right) _i\left( \vec k\times \vec k^{\prime }\right) _{i^{\prime
}}e^{i\left( \vec k+\vec k^{\prime }\right) .\left( \vec y-\vec y^{\prime
}\right) }\delta G_{1k}(y_0,y_0^{\prime })\delta G_{1k^{\prime
}}(y_0,y_0^{\prime })  \label{ap}
\end{equation}
where 
\begin{equation}
G_{1k}^0=G_k^{+0}+G_k^{-0}=\frac{\cos \Omega _k(y_0,y_0^{\prime })}{\sqrt{%
\omega _k(y_0)\omega _k(y_0^{\prime })}}  \label{aq}
\end{equation}
with $\Omega _\kappa (y_0,y_0^{\prime })=\Omega _\kappa (y_0)-\Omega _\kappa
(y_0^{\prime })$, 
\begin{equation}
\delta G_{1k}(y_0,y_0^{\prime })=2\alpha _k\beta _k^{*}f_k^0(y_0)f_\kappa
^0(y_0^{\prime })+2\alpha _k^{*}\beta _kf_k^{0*}(y_0)f_\kappa
^{0*}(y_0^{\prime })+2\left| \beta _k\right| ^2G_{1k}^0(y_0,y_0^{\prime })
\label{ar}
\end{equation}
The energy density of the magnetic field can be calculated from the two
point correlation function generated by the stochastic current: 
\begin{eqnarray}
\left\langle B(\vec x)B(\vec x^{\prime })\right\rangle &=&H^4\int \frac{dy_0d%
\vec y}{\left( 2\pi \right) ^2}\frac{dy_0^{\prime }d\vec y^{\prime }}{\left(
2\pi \right) ^2}G^{ret}(y_0,\vec y;x_0,\vec x)G^{ret}(y_0^{\prime },\vec y%
^{\prime };x_0,\vec x^{\prime })\times  \label{at} \\
&&\ \ \ \times \left\{ \bar N_{ii}^1\left[ \left( y_0,\vec y\right) ,\left(
y_0^{\prime },\vec y^{\prime }\right) \right] +\bar N_{ii}^2\left[ \left(
y_0,\vec y\right) ,\left( y_0^{\prime },\vec y^{\prime }\right) \right]
\right\}  \nonumber
\end{eqnarray}
Where $G^{ret}$ are the causal propagators for the Maxwell field, obtained
from the homogeneous solutions to Eq. (\ref{ad}). By transforming Fourier
the causal propagators, the spatial integrals can be readily performed by
virtue of the simple spatial dependence of the propagators. We have 
\begin{eqnarray}
\left\langle B(\vec x)B(\vec x^{\prime })\right\rangle &=&H^4\int
dy_0dy_0^{\prime }\times \int \frac{d\vec k\,d\vec k^{\prime }}{\left( 2\pi
\right) ^3}\left| \vec k\times \vec k^{\prime }\right| ^2\times \left\{ 4q^2%
\bar G_{1k}^0(y_0,y_0^{\prime })\delta \bar G_{1k^{\prime }}(y_0,y_0^{\prime
})+\right.  \label{au} \\
&&\ +\left. q^2\delta \bar G_{1k}(y_0,y_0^{\prime })\delta \bar G%
_{1k^{\prime }}(y_0,y_0^{\prime })\right\} e^{i\left( \vec k+\vec k^{\prime
}\right) .\left( \vec x-\vec x^{\prime }\right) }G_{\left| k+k^{\prime
}\right| }^{ret}(y_0,x_0)G_{\left| k+k^{\prime }\right| }^{ret}(y_0^{\prime
},x_0)  \nonumber
\end{eqnarray}

The energy density of the magnetic field coherent over a given dimensionless
spatial scale $\lambda $ is given by $E_\lambda =\frac 1{a^4V_\lambda ^2}%
\int d\vec x\int d\vec x^{\prime }\left\langle B(\vec x)B(\vec x^{\prime
})\right\rangle $ which amounts to insert the window function ${\cal W}%
_{kk^{\prime }}(\lambda )\equiv V_\lambda ^{-1}\int_{V_\lambda }d\vec x%
\;e^{i\left( \vec k+\vec k^{\prime }\right) .\vec x}$. The energy density
reads $\rho _B=a^{-4}\left\langle B_\lambda ^2\right\rangle $, where 
\begin{eqnarray}
\left\langle B_\lambda ^2\right\rangle &=&H^4\int dy_0dy_0^{\prime }\times
\int \frac{d\vec k\,d\vec k^{\prime }}{\left( 2\pi \right) ^3}{\cal W}%
_{kk^{\prime }}^2(\lambda )\left| \vec k\times \vec k^{\prime }\right|
^2G_{\left| k+k^{\prime }\right| }^{ret}(y_0,x_0)G_{\left| k+k^{\prime
}\right| }^{ret}(y_0^{\prime },x_0)  \label{av} \\
&&\ \ \ \ \ \ \ \ \ \times \left\{ 4q^2\bar G_{1k}^0(y_0,y_0^{\prime
})\delta \bar G_{1k^{\prime }}(y_0,y_0^{\prime })+q^2\delta \bar G%
_{1k}(y_0,y_0^{\prime })\delta \bar G_{1k^{\prime }}(y_0,y_0^{\prime
})\right\}  \nonumber
\end{eqnarray}
As the window function satisfies ${\cal W}_{kk^{\prime }}(\lambda )\approx 1$
for $\left| \vec k+\vec k^{\prime }\right| \lambda \lesssim 1$ and ${\cal W}%
_{kk^{\prime }}(\lambda )\approx 0$ for $\left| \vec k+\vec k^{\prime
}\right| \lambda \gtrsim 1$, we can take as the upper limit in the momentum
integrals, $k_{\max }\simeq \frac 1\lambda $ and ${\cal W}_{kk^{\prime }}=1$
in the interval $(0,k_{\max })$. All the cosmological interesting scales are
such that $k\ll 1$: For example for a galaxy we have that its physical scale
today is (if it had not collapsed) $\lambda _G\simeq 1Mpc\simeq
10^{38}GeV^{-1}=\lambda _cT_{rh}/T_{today}\simeq 10^{28}\lambda _c$ ($%
T_{rh}\simeq 10^{15}GeV$ is the temperature of the Universe at the end of
reheating and $T_{today}\simeq 10^{-13}GeV$ is the present temperature of
the microwave background). The commoving wavelength is $\lambda _c\simeq
10^{10}GeV^{-1}$, taking $H=10^{11}GeV$ the dimensionless scale $\lambda
=H\lambda _c\simeq 10^{21}$ and the corresponding dimensionless momentum $%
k\simeq 10^{-21}$.

The retarded propagators are constructed with the homogeneous solutions of
Eq. (\ref{ad}). If we take the conductivity (cfr. Refs. \cite{widrow,olinto}%
) $\sigma \simeq Te^{-2}=T_{rh}e^{-2}(1+y_0)^{-1}=\sigma _0$ $(1+y_0)^{-1}$, 
$\sigma _0\simeq e^{-2}T_{rh}=e^{-2}\sqrt{Hm_{pl}}=He^{-2}\sqrt{m_{pl}/H}$, $%
\bar \sigma _0\equiv \sigma _0/H=e^{-2}\sqrt{m_{pl}/H}$ being the
dimensionless conductivity, the homogeneous equation (\ref{ad}) reads (for
scales larger than the horizon) 
\begin{equation}
\frac{\partial ^2}{\partial y_0^2}B_k+\frac{\bar \sigma _0}{(1+y_0)}\frac 
\partial {\partial y_0}B_k=0  \label{aw}
\end{equation}
where we have used $B_k(y_0)=\int d^3ye^{-i\vec k.\vec y}B(\vec y,y_0)$. The
causal propagator for this equation is given by 
\begin{equation}
G_{\left| k+k^{\prime }\right| }^{ret}=-\frac{1+y_0}{\bar \sigma _0-1}\left[
1-\left( \frac{1+y_0}{1+x_0}\right) ^{\bar \sigma _0-1}\right]  \label{aww}
\end{equation}
If $x_0\gg y_0$ the final expression for the propagator is $G_{\left|
k+k^{\prime }\right| }^{ret}\simeq (1+y_0)/\bar \sigma _0$.

In spite of the fact that the scalar field is not in thermal equilibrium
with the background radiation, the electromagnetic interaction with it
causes a correction to the value of the field mass, proportional to the
temperature of the bath, i.e. we have $%
m^2=m_0^2+eT^2=m_0^2+eT_{rh}^2/(1+y_0)^2$. We will consider this correction
in the frequencies $\omega _k(y_0)$ but not in the Bogolubov coefficients
because we assume instantaneous particle creation. In fact, when the
transition from Inflation to radiation dominance occurs, large numbers of of
particles are created very quickly by parametric resonance. Thermalization
is a process that takes place afterwards and during a longer period of time 
\cite{boya}. As for the ratio $m/H,$ if we consider, for example, the Higgs
mass ($m\simeq 10^2GeV$), we have $m/H\simeq 10^{-9}$. We can therefore
neglect the terms $k^2$ in front of those $m^2/H^2$. When we perform the
products $\bar G_{1k}^0(y_0,y_0^{\prime })\delta \bar G_{1k^{\prime
}}(y_0,y_0^{\prime })$ and $\delta \bar G_{1k}(y_0,y_0^{\prime })\delta \bar 
G_{1k^{\prime }}(y_0,y_0^{\prime })$ in Eq. (\ref{av}) we will have terms
where the exponentials cancel and terms where this does not occur. The
latter oscillate and hence can be neglected because they will give a smaller
contribution than the first ones when the time integrals are performed.
These integrals are performed between $0$ and $\Upsilon $, where $\Upsilon $
is the dimensionless time when a scale $k^{-1}$ reenters the horizon; after
this time, the current quickly relaxes to its equilibrium value through
particle - antiparticle annihilation. From the expression of the Bogolubov
coefficients, Eqs. (\ref{am}) we can see that the contribution from the term
quartic in $\beta _k$ overwhelms the quadratic one. The leading contribution
to (\ref{av}) reads 
\begin{equation}
\left\langle B_\lambda ^2\right\rangle \sim \frac 43q^2\left( \frac H{m_0}%
\right) ^6H^4k_{\max }^4\left( \frac 1{\bar \sigma _0}\left. \sqrt{\frac{%
m_0^2\left( 1+y_0\right) ^2}{H^2}+\frac{eT_{rh}^2}{H^2}}\right| _0^\Upsilon
\right) ^2  \label{be}
\end{equation}
For the scale considered, $k_{\max }\simeq 10^{-21}$, for the electroweak
mass scale, $m_0\simeq 10^2GeV$, $\bar \sigma _0\simeq e^{-2}10^4$, $%
T_{rh}\simeq 10^{15}GeV$, $q^2=e^2=1/137$ and recalling $\Upsilon \sim \pi
/\left( 2k_{\max }\right) $ we first note that, by simply replacing these
figures in Eq. (\ref{be}), the term proportional to $m_0^2\left(
1+y_0\right) ^2$ overwhelms the one proportional to $T_{rh}^2$ in the square
root by a factor $\sim 10^{20}$. Therefore we can take 
\begin{equation}
\left\langle B_\lambda ^2\right\rangle \sim \frac 43q^2\left( \frac H{m_0}%
\right) ^6H^4k_{\max }^4\frac 1{\bar \sigma _0^2}\frac{m_0^2\Upsilon ^2}{H^2}%
=\frac 43q^2\left( \frac H{m_0}\right) ^4H^4k_{\max }^2\frac 1{\bar \sigma
_0^2}  \label{bee}
\end{equation}
\begin{equation}
\left\langle B_\lambda ^2\right\rangle \sim 10^{24}GeV^4\rightarrow r\equiv 
\frac{\rho _B}{\rho _{bck}}\sim 10^{-36}  \label{bf}
\end{equation}
where $\rho _{bck}=T_{rh}^4$ is the energy density of the CMBR. These values
correspond to a commoving field of strength $B_\lambda \sim 10^{32}Gauss.$
The physical field $B_{ph}=a^{-2}B_\lambda $, such that $\rho _B\equiv $ $%
B_{ph}^2$, gives a present value of 
\begin{equation}
B_{ph}\sim 10^{-24}Gauss  \label{bg}
\end{equation}
This value satisfies the constraints imposed by the anisotropy in the
microwave background \cite{barrow} and Big Bang nucleosynthesis \cite{bbn}
and (marginally) suffices to seed the galactic dynamo \cite{zeldovich,olinto}
, \cite{dimop}. Stronger fields are obtained by diminishing the value of the
field mass. We can estimate the lower bound to the mass by demanding that
today $B_{ph}\sim 10^{-9}Gauss$, according to the mentioned constraints. We
obtain $m\gtrsim 10^{-6}GeV$. Considering the value of the mass used
throughout the paper as the upper bound we find the range $%
10^{-6}GeV\lesssim m\lesssim 10^2GeV$. Fields outside of this range will
contradict observational and/or theoretical constraints.

\section{Conclusions}

To conclude, we have presented a new mechanism for the generation of a
primordial magnetic field, based on the breaking of conformal invariance. In
previous works \cite{widrow,dolgov,ratra}, conformal invariance was broken
in order to amplify quantum fluctuations of $\vec B$. The amplification was
in general very small. Here scalar, massive charged particles minimally
coupled to gravity, and coupled to the electromagnetic field produce
stochastic fluctuations in the source of Maxwell eqs. which in turn generate
an astrophysically relevant primordial field. For conformally coupled
massive scalar field, the Bogolubov coefficient scale with the momenta as $%
k^{-1/2}$ which means that the number of created particles is very small and
consequently the magnetic field generated will be too weak to be
astrophysically interesting ($\sim 10^{-80}$ times smaller than in the
minimally coupled case, for the physical parameters used in this paper). The
fact that spinor fields are conformal invariant if massless, manifests
itself in that the proposed mechanism does not generate strong enough
magnetic fields, as is shown in the Appendix. Finally we must say that the
value of the field quoted in Eq. (\ref{bg}) is to be considered as a rough
estimate; a more precise evaluation will require a more detailed
consideration of the whole electroweak gauge theory and of the
non-equilibrium evolution of current and field.

\section{Appendix}

Here we show that the magnetic field generated by an electric current
originated by the creation of fermionic particles is indeed negligible. The
Dirac equation for spinors in a FRW Universe reads (in conformal time) \cite
{claudio,silvia} 
\begin{equation}
\left[ i\gamma ^\mu \partial _\mu -ma(y_0)\right] \chi (y_0,\vec y)=0
\label{ffa}
\end{equation}
where $\chi (y_0,\vec y)=a^{3/2}(y_0)\psi (y_0,\vec y)$. $y_0$ and $\vec y$
are the dimensionless time and space coordinates defined in the paper and in
this Appendix $m$ is to be understood as a dimensionless mass, i.e. $%
m\rightarrow m/H$. The electric current reads 
\[
j^\mu =e\bar \chi (y_0,\vec y)\gamma ^\mu \chi (y_0,\vec y) 
\]
where $\bar \chi (y_0,\vec y)=\chi ^{\dagger }(y_0,\vec y)\gamma ^0$, i.e.
the Dirac adjoint. The positive and negative energy spinors, $u_{ks}$ and $%
v_{ks}$ read 
\begin{eqnarray}
u_{ks} &=&\left( 
\begin{array}{c}
\frac{\vec \sigma .\vec k}{k^2}\left( ma(y_0)+i\frac{\dot f_k(y_0)}{f_k(y_0)}%
\right) C_s \\ 
C_s
\end{array}
\right) f_k(y_0)e^{i\vec k.\vec y}  \label{fa} \\
v_{ks} &=&\left( 
\begin{array}{c}
C_s \\ 
-\frac{\vec \sigma .\vec k}{k^2}\left( ma(y_0)-i\frac{\dot f_k^{*}(y_0)}{%
f_k^{*}(y_0)}\right) C_s
\end{array}
\right) f_k^{*}(y_0)e^{-i\vec k.\vec y}  \label{fb}
\end{eqnarray}
where $f_k(y_0)$ is a solution to the equation 
\begin{equation}
\left[ \frac{\partial ^2}{\partial y_0^2}+k^2+m^2a^2(y_0)-im\frac{\partial
a(y_0)}{\partial y_0}\right] f_k=0  \label{fc}
\end{equation}
The spinors are normalized according to the (time independent) product $%
\int_\Sigma d\Sigma \bar \chi \gamma ^0\chi $, $\Sigma $ being a spacelike
hypersurface.

The noise kernel (\ref{ag}) due to this field reads (after a rather long
calculation) 
\begin{eqnarray}
N_{ii}(y_0,y_0^{\prime }) &=&4q^2\int d\vec k\int d\vec p\frac{%
f_k(y_0^{\prime })f_p(y_0^{\prime })f_k^{*}(y_0)f_p^{*}(y_0)}{N_kN_pN_kN_p}%
\times  \label{fd} \\
&&\ \ \ \ \ \ \left\{ {\cal M}_{k,p}^{(1)}(y_0,y_0^{\prime })+{\cal M}%
_{k,p}^{(2)}(y_0,y_0^{\prime })+\left| \vec k+\vec p\right| ^2\right\} 
\nonumber
\end{eqnarray}
where 
\begin{eqnarray}
{\cal M}_{k,p}^{(1)}(y_0,y_0^{\prime }) &=&\frac{\left| \vec k+\vec p\right|
^2}{k^2p^2}\left[ ma(y_0^{\prime })+i\frac{\dot f_k(y_0^{\prime })}{%
f_k(y_0^{\prime })}\right] \left[ ma(y_0^{\prime })+i\frac{\dot f%
_p(y_0^{\prime })}{f_p(y_0^{\prime })}\right]  \label{fe} \\
&&\ \ \ \ \ \ \times \left[ ma(y_0)-i\frac{\dot f_k^{*}(y_0)}{f_k^{*}(y_0)}%
\right] \left[ ma(y_0)-i\frac{\dot f_p^{*}(y_0)}{f_p^{*}(y_0)}\right] 
\nonumber
\end{eqnarray}
\begin{eqnarray}
{\cal M}_{k,p}^{(2)}(y_0,y_0^{\prime }) &=&\frac{\left[ \vec k.\left( \vec k+%
\vec p\right) \right] \left[ \left( \vec k+\vec p\right) .\vec p\right] }{%
k^2p^2}\left\{ \left[ ma(y_0^{\prime })+i\frac{\dot f_k(y_0^{\prime })}{%
f_k(y_0^{\prime })}\right] \left[ ma(y_0^{\prime })+i\frac{\dot f%
_p(y_0^{\prime })}{f_p(y_0^{\prime })}\right] +\right.  \label{ff} \\
&&\ \ \ \ \ \ +\left. \left[ ma(y_0)-i\frac{\dot f_k^{*}(y_0)}{f_k^{*}(y_0)}%
\right] \left[ ma(y_0)-i\frac{\dot f_p^{*}(y_0)}{f_p^{*}(y_0)}\right]
\right\}  \nonumber
\end{eqnarray}
$N_{k,p}$ are the normalization factors which we calculate for Inflation.
The solution to the field equation (\ref{fc}) for the inflationary period
that corresponds to positive frequency for $y_0\rightarrow -\infty $ read 
\begin{equation}
f_k(y_0)=\left( 1-y_0\right) ^{1/2}H_\nu ^{(1)}\left[ k\left( 1-y_0\right)
\right] ;\quad \nu =\frac 12+im  \label{fg}
\end{equation}
and therefore 
\begin{equation}
N_k=N_k=\frac 2{\sqrt{\pi }}\frac 1{k^{1/2}}  \label{fh}
\end{equation}
For the radiation dominated period and for $k/m\ll 1$ we write the mode
functions as \cite{abramowitz} 
\begin{equation}
f_k=\alpha _ke^{im(1+y_0)^2/2}+\beta _ke^{im(1+y_0)^2/2}\int_{z(y_0)}^\infty
e^{-s^2/2}ds  \label{fi}
\end{equation}
where $z(y_0)=(1+i)\sqrt{m}(1+y_0)$ and $\alpha _k$ and $\beta _k$ are the
Bogolubov coefficients obtained by matching the modes and its time
derivatives at $y_0=0$. In the limit $k/m\ll 1$ they read 
\begin{eqnarray}
\alpha _k &=&-\frac i\pi \Gamma \left( \frac 12+im\right) e^{im/2}\left( 
\frac 2k\right) ^{1/2+im}  \label{fj} \\
\beta _k &=&-\frac i{2\pi }\Gamma \left( -\frac 12+im\right) e^{im/2}\left( 
\frac 2k\right) ^{1/2+im}k^2  \label{fk}
\end{eqnarray}
We see that in this case there is no divergence in the number of quanta
created and therefore the expected electric current will be very small. We
replace the modes (\ref{fi}) in the equations (\ref{fe}) and (\ref{ff}) and
with the obtained expressions we evaluate 
\begin{equation}
\left\langle B^2\right\rangle =H^4\int \frac{dy_0}{\left( 2\pi \right) ^2}%
\frac{dy_0^{\prime }}{\left( 2\pi \right) ^2}G^{ret}(y_0,x_0)G^{ret}(y_0^{%
\prime },x_0)N_{ii}(y_0,y_0^{\prime }){\cal W}_{kp}^2(\lambda )  \label{fl}
\end{equation}
with the same considerations that were made for the scalar field. The main
contribution comes from the first two terms and they are respectively 
\begin{eqnarray}
\ \left\langle B^2\right\rangle ^{(1)} &\simeq &q^2H^4\frac{%
8e^{2im(1+T_{\max })^2}}{m^2\bar \sigma _0^2\left[ \frac 14+m^2\right]
^2\cosh ^2\left[ \pi m\right] }k_{\max }^{10}  \label{fm} \\
\ \left\langle B^2\right\rangle ^{(2)} &\simeq &q^2H^4\frac{64}{m^2\bar 
\sigma _0^2\left[ \frac 14+m^2\right] \cosh ^2\left[ \pi m\right] }k_{\max
}^8  \label{fn}
\end{eqnarray}
For example, we have that for a mass $m=10^{-11}$ (i.e. physical mass $\sim
1 $ $GeV$ in units of $H=10^{11}$ $GeV$), $k_{\max }$ and $\bar \sigma _0$
the same as the ones used in the paper, $\left\langle B^2\right\rangle
^{(1)}\sim 10^{-160}GeV^4$ and $\left\langle B^2\right\rangle ^{(2)}\sim
10^{-116}GeV^4$, values completely negligible in comparison with the one
quoted in Eq. (\ref{bf}).

\section{Acknowledgments}

We are indebted to F. M. Spedalieri and F. H. Gaioli for enlightening
discussions.This work was partially supported by Universidad de Buenos
Aires, CONICET, Fundaci\'on Antorchas, and by the Comission of the European
Communities under contract Nr. C11*-CJ94-0004.

\end{document}